\documentclass[preprint2]{aastex}

\shorttitle{Disk stability under MONDian gravity}
\shortauthors{M. A. Jim\'enez and X. Hernandez}

\begin{document}


\title{Local galactic disk stability under MONDian gravity}


\author{M. A. Jim\'enez and X. Hernandez}
\affil{Instituto de Astronom\'{\i}a, Universidad Nacional Aut\'{o}noma de M\'{e}xico,
Apartado Postal 70--264 C.P. 04510 M\'exico D.F. M\'exico}


\altaffiltext{2}{email:xavier@astro.unam.mx}


\begin{abstract}
Toomre's $Q$ stability parameter has long been shown through various theoretical arguments and numerical simulations,
to be the principal determinant of stability against self-gravity in a galactic disk, under classical gravity. Comparison 
with observations however, has not always confirmed the condition of $Q=Q_{crit}$ to be well correlated with various critical 
galactic radii. In this paper we derive the analogous critical parameter, $Q_{M}$, under MONDian gravity. The result is a 
modification by a factor of $(\sigma \Omega)/a_{0}$, $Q_{M}=(\sigma \Omega)^{2}/(a_{0}G \Sigma)$, where $a_{0}$ is the critical 
acceleration scale of MOND. We then show through a direct comparison to a homogeneous sample of observed disk galaxies with 
measured brightness profiles, rotation velocity curves and internal velocity dispersion profiles, that the critical radii at 
which brightness profiles dip below the exponential fit, are much more accurately predicted by $Q_{M}$ than by the $Q$ of 
classical gravity. This provides a new and completely independent argument supporting the reality of a change in the form of 
gravity on reaching the low acceleration regime.
\end{abstract}



\section{Introduction}

Toomre's $Q$ parameter (Toomre 1964) neatly encompasses a comparison of the disruptive total tidal forces and internal
dynamical pressure, against local self gravity, for a local element of a galactic disk. It remains the most basic
measure of the stability of a galactic disk, a valuable diagnostic in evaluating the propensity of a disk towards
spiral arm, bar or bulge formation e.g. Athanassoula (2003). Also, $Q$ has often been related to star formation processes 
in spiral galaxies. The idea being that in very hot disk regions, above certain critical $Q$ values, the self gravity of a
gaseous element is insufficient to overcome the combined effects of internal pressure and tidal shears, and is hence
incapable of initiating collapse and undergoing star formation. The expectation then arises for surface brightness
profiles to show truncations or kinks at the critical radii where $Q$ crosses some threshold value. 

In terms
of an evolutionary view of a galactic disk, all the parameters upon which $Q$ depend, the rotation curve, the 
surface density profile and the internal velocity dispersion one, are expected to change over the course of
a disk galaxy's growth and lifetime e.g. Firmani et al. (1996). Thus, the present day surface brightness profile,
an integral over the galactic lifetime of the star formation processes, will retain only a somewhat averaged
out version of the history of this critical radius and its evolution e.g. Schaye (2004). However, observational
studies seeking to relate features in the surface density profiles of disks to estimates of $Q$, have often
found only weak correlations with critical $Q$ values e.g. the subcritical large $Q$ star-forming disks of 
Martin \& Kennicutt (2001).

In view of the above, we find it interesting to develop the analogous stability parameter under MONDian gravity, 
$Q_{M}$, and to perform a first observational test of this new parameter. MONDian gravity, where the force
felt by a test particle orbiting a mass $M$ changes from the Newtonian value to $(a_{0}GM)^{1/2}/R$, was originally
presented as a way of understanding the observed asymptotic flatness of galactic rotation curves in the absence
of any dark matter, Milgrom (1983). Since, modified gravity schemes that reduce to the above MONDian behaviour, 
have been shown to be accurately in accordance with a range of gravitational astrophysical observations and scales;
in Hernandez et al. (2012) some of us showed solar neighbourhood wide binary stars to have relative velocities which
deviate from Keplerian expectation, precisely on crossing the $a_{0}$ threshold, Haghi et al. (2011), 
Scarpa et al. (2011) and Hernandez et al. (2013a) showed that the same transition is observed in Galactic globular 
clusters, which show consistency also with a galactic Tully-Fisher relation, Kroupa (2012) has given an accurate
description of the dynamics and orbits of dwarf galaxies, and extending the well known description of Spiral
rotation curves under MOND (e.g. Sanders \& McGaugh 2002, Swaters et al. 2010), in Jimenez et al. (2013) some of us
presented a full description of the massive elliptical galaxy NGC 4649, all of the above cases, without the substantial 
fractions of dark matter which a classical description requires.

In the above, the range of gravitational phenomena studied under MONDian gravity has been progressively extended from
equilibrium rotation velocities, to an ever growing range of aspects. This provides increasingly more independent
empirical tests, e.g. the isothermal distributions of MONDian gravity which accurately and naturally reproduce the 
observed $\rho\propto r^{-3}$ density profiles of tenuous stellar halos surrounding external galaxies, 
Hernandez et al. (2013b). In this paper we continue along the same lines, by developing a MONDian analogue to the
standard first order Toomre's stability criterion for disk galaxies, and preform a first test using the recent and
homogeneous galactic structure atlas of Martinsson et al. (2013a) and Martinsson et al. (2013b). 

Global disk stability 
under MOND has been treated by Milgrom (1989), however, within the limits of tightly wound global, purely surface density 
perturbations, imposing an infinitely thin disk. This last constraint, one of treating an essentially 3D problem within 
a 2D approximation, leads to the modified Toomre criterion of Milgrom (1989) where only an order unity correction with 
respect to the classic case appears through the dimensionless MOND transition function, and where the critical acceleration 
of $a_{0}$ does not appear.

Section (2) presents a first order local 3D derivation of Toomre's $Q$ stability parameter under MONDian gravity, $Q_{M}$. 
In section (3) we use a sample of disk galaxies from Martinsson et al. (2013a) and Martinsson et al. (2013b) with 
measured surface density light profiles, rotation velocities and internal disk dispersion velocities, to test the 
relevance of $Q_{M}$ in terms of predicting critical radii at which the surface brightness profile falls below the 
exponential fit. We see from the data that indeed, $Q_{M}$ offers a much better description of the observed break radii 
than $Q$. Our conclusions are presented in section(4).

\section{Theoretical expectations}

In the study of classical gravitational instabilities in galactic disks, the Toomre stability criterion
is regarded as the most valuable diagnostic. This is defined as:

\begin{equation}
Q_{T}=\frac{\sigma \kappa}{\alpha G \Sigma},
\end{equation}

In the above, $\kappa(R)$ is the epicyclic frequency of the disc, related to the orbital frequency in the disk, 
$\Omega(R)$, through $\kappa^{2}(R) = R d\Omega^{2} /dR + 4 \Omega^{2}$, $\sigma(R)$ represents the velocity dispersion of 
disk material, $\Sigma(R)$ the stellar surface density and $\alpha$ is a numerical constant of order one which varies 
slightly for the cases of non-collisional stellar components, dissipative gaseous disks, mixed components, the presence 
of magnetised gas, etc. Binney \& Tremaine (1987). The above stability criterion arises from the dispersion relation for 
infinitely thin disks (Toomre 1969), but has been shown through extensive numerical simulations to be the most relevant
first order stability determinant for realistic galactic disks, when assuming standard gravity.

Under Newtonian gravity one can understand the basic physics involved in Toomre's stability criterion by considering a 
parcel of disk material of density $\rho$ moving at the orbital frequency of the disk at radius $R$, $\Omega(R)$. 
Such a parcel will be prevented from collapsing onto itself due to its self gravity by the overall tidal forces, if its 
density is roughly below that of the total matter density interior to $R$, $\rho<\bar{\rho}(R)$, i.e. it fails to satisfy 
a tidal density criterion. To within a numerical constant of order one, $\Omega^{2}(R)=G\bar{\rho}(R)$, so we can write the
tidal stability criterion:

\begin{equation}
\rho < \frac{\Omega^{2}}{G}
\end{equation}

Taking $\Sigma= \rho h$, with $h$ the typical scale height of the disk, yields

\begin{equation}
h_{t} > \frac{G \Sigma}{\Omega^{2}}
\end{equation}

\noindent as the critical disk scale above which tides will stabilise local self-gravity.

On the other hand, internal pressure will stabilise the disk element against its self-gravity, below a critical Jeans 
scale, $h_{J}=\sigma/(G \rho)^{1/2}$. Taking again $\Sigma= \rho h$ gives:

\begin{equation}
h_{J} < \frac{\sigma^{2}}{G \Sigma}
\end{equation}

\noindent as the critical disk scale below which internal pressure will stabilise local self-gravity.

The disk will be locally stable against perturbations of all scales below a critical $\Sigma$ such that
$h_{J}=h_{t}$, as below this critical $\Sigma$, $h_{J}>h_{t}$, with the equalling of the two critical scales above
giving the condition:

\begin{equation}
Q=\frac{\sigma \Omega}{G \Sigma}=1
\end{equation}

\noindent as the critical criterion. For $Q>1$ the disk is locally stable against perturbations of all scales.
In eq.(5) above, we see the numerator representing dynamical pressure and tides through $\sigma$ and $\Omega$
as stabilising factors against the local self-gravity in the denominator through $G \Sigma$.

We see that this last expression neatly captures the essential physics behind $Q_{T}=1$ of eq.(1) being the relevant
disk stability criterion, specially considering that this critical radius is expected to occur within the flat rotation
curve region of a galactic disk, where $\kappa=2^{1/2} \Omega$. For a canonical value of $\alpha=\pi$ in eq.(1), the critical
point would appear as $(0.45\sigma \Omega /G \Sigma) =1$ within the flat rotation curve region. The above development 
is well known (e.g. Binney \& Tremaine 1987), but is repeated here to make the analogy with the development under modified 
gravity which follows more explicit. 

In going to a MONDian modified gravity regime, for accelerations below a critical value of $a_{0}$, the force felt by a
test particle orbiting a spherically symmetric mass distribution of total mass M becomes $F_{M}=(a_{0} GM)^{1/2}/R$,  
and hence rotation velocities become flat at a value consistent with the observed Tully-Fisher relation, $V_{f}=(a_{0} GM)^{1/4}$.
It is easy to see that the resulting tidal criterion for a spherically symmetric mass element now becomes 
$\rho< (h_{tM}/R) \bar{\rho}(R)$ e.g. Hernandez \& Jimenez (2012). Since the critical point will occur at radii such
that the integrated mass of the galactic disc has essentially converged, $V_{f}^{4}=a_{0} G \bar{\rho}(R) R^{3}$. Using
again $\Sigma= \rho h$ to eliminate $\rho$ in favour of the $\Sigma$ of eq.(5), the critical scale above which tides will 
stabilise local self-gravity becomes:

\begin{equation}
h_{tM}>\frac{(a_{0}G \Sigma)^{1/2}}{\Omega^{2}}
\end{equation}

The modified version of the Jeans scale, the balance between internal dynamical pressure and self-gravity is now
$m_{JM}=\sigma^{4}/Ga_{0}$, with parcels of mass below $m_{JM}$ being stabilised by their internal dynamical 
pressure, e.g. Mendoza et al. (2011). Taking $m_{JM}=\Sigma h_{JM}^{2}$ to replace $m_{JM}$ for the $\Sigma$ of eq.(5) 
leads to the condition:

\begin{equation}
h_{JM} < \frac{\sigma^{2}}{(a_{0} G \Sigma)^{1/2}}
\end{equation}

\noindent Scales below $h_{JM}$ will be stabilised by internal velocity dispersion. Following the analogy with the
Newtonian derivation of eq.(5), we see that under the modified regime, the disk will be locally stable against perturbations 
of all scales below a critical $\Sigma$ such that $h_{JM}=h_{tM}$, as below this critical $\Sigma$, $h_{JM}>h_{tM}$,
with the equalling of the two critical scales above giving the condition:

\begin{equation}
Q_{M}=\frac{(\sigma \Omega)^{2}}{a_{0} G \Sigma} =1
\end{equation}

\noindent as the critical criterion under modified gravity. For $Q_{M}>1$ the disk is locally stable against perturbations 
of all scales. Introducing the characteristic surface density of MOND, $\Sigma_{M}=a_{0}/G$, we can write the critical surface
density which yields $Q_{M}=1$ as:

\begin{equation}
\Sigma_{c}=\Sigma_{M}\left( \frac{\sigma \Omega}{a_{0}}  \right)^{2}
\end{equation}

We see that the result of eq.(8) introduces a dimensionless factor of $(\sigma \Omega/a_{0})$ to the
classical expression of eq.(5) i.e., $Q_{M}=(\sigma \Omega/a_{0}) Q$. This correction factor will be large when the 
internal disk acceleration scale is much smaller than $a_{0}$. Thus, we see that a galactic disk which might appear as
``sub-critical'' in terms of surface density because its $Q>1$, could quite easily have a low $Q_{M}<1$ 
($\sigma \Omega/a_{0} <1$) and hence be naturally understood as hosting the observed levels of star formation.
For values of $\Sigma>\Sigma_{M}$, the classical $Q$ would apply.

It is interesting that under classical gravity stability to both global tight winding patterns under a 2D treatment, and
a local bulk self-gravity vs. dynamical pressure and tides balance give essentially the same result of $Q$, while for MONDian
gravity, the former leads to the modified $Q$ of Milgrom (1989) where only a dimensionless correction factor given by the MOND
transition function appears, and the latter to the $Q_{M}$ of eq.(8). Notice also that stellar disk critical radii occur in 
the flat rotation curve region, and at surface densities well below the critical MOND value of $a_{0}/G$ (e,g, Famaey \& McGaugh 2012), 
making the preceding development valid not only for a MONDian gravity scheme where no external field effect is included, but 
also under MOND as such.

We can reach a slightly more approximate, but more easily testable prediction by taking the following empirically grounded 
models for $\Sigma(R)$ and $\sigma(R)$:
  
\begin{equation}
\Sigma(R)= \Sigma_{0} e^{-(R/R_{\star})} 
\end{equation}

\begin{equation}
\sigma(R)= \sigma_{0} e^{-(R/R_{\sigma})} 
\end{equation}

Taking the common observational result of $R_{\sigma} \approx 2R_{\star}$, if we now evaluate $Q_{M}$ at $R_{c}$, 
the critical galactic radius, where $Q_{M}=1$, we obtain the relation:

\begin{equation}
\frac{\sigma^{2}_{0} \Omega^{2}(R_{c})}{a_{0} G \Sigma_{0}}=1
\end{equation}

Multiplying above and below by $R^{2}_{\star}$, and using the relation $M=2\pi R^{2}_{\star} \Sigma_{0}$ for 
galactic discs where again $a_{0}GM=V_{f}^{4}$, with $V_{f}$ is the flat rotation curve disk orbital velocity yields,

\begin{equation}
\frac{R_{c}}{R_{\star}}  = \beta \frac{\sigma_{0}}{V_{f}},
\end{equation}  

\noindent where $\beta$ is a proportionality constant to account for the various first order estimates introduced into the 
previous development. Thus, we reach an easily testable prediction for the critical disk radii, in units of the stellar disk 
scale length, to be proportional to the quotient of the central disk dispersion velocity to the flat rotation velocity value, 
if MONDian dynamics are important to real galactic disks.

\section{Comparisons with real galaxies}

We begin with a first test using only eq.(13), where use of eq.(10) and eq.(11) allows to eliminate the surface mass density of the 
disk, a parameter which is subject to significant uncertainties due to the large error intervals in mass to light ratios, gas fractions, 
etc. The galactic parameters needed for testing the relevance of the stability criteria for galactic disks under MONDian gravity of 
eq.(13) are: the value of the rotation velocity within the flat rotation curve region $V_f$, the $\sigma_{0}$ parameter from 
a full disk fit of eq. (11) to the velocity dispersion profile, and $R_{\star}$ and $R_c$, the disk scale, and the critical
galactic radius where the brightness surface profile dips below the exponential fit. Notice that $\sigma_{0}$ is not the actual 
$\sigma(R=0)$, which would mostly reflect central bulge dynamics, and not the velocity dispersion representative of the disk at large 
radii through eq.(11).

Fortunately, all these parameters appear in the DiskMass Survey of Martinsson et al. (2013a), which consists of a sample of thirty 
disk galaxies, with their corresponding rotation curves, the velocity dispersion profiles of the disk stars measured along the line of 
sight and perpendicular to the disk plane, the  K- band surface brightness profiles, and observed disk scale lengths for each galaxy in 
the sample. We prefer to work with a homogeneous sample where all the quantities for all the galaxies have been measured and treated 
consistently by the same authors, rather than attempt to use a larger heterogeneous sample where any trend will always be suspect to 
having arisen through the shifting systematics across the sample.

To test eq.(13) we use observational the quantities $V_{f}$ and $R_{\star}$ from Martinsson et al.(2013a) and Martinsson et al.(2013b), 
$R_c$ is obtained directly from K-band surface brightness profile, we take $\sigma_{0}$ from the exponential fits to the velocity 
dispersion profile presented in the same work. In fact we use the fit for $\sigma_{z}$ and use the parameter $\sigma_{z,0}$ which is 
not affected by the varying degrees of inclination at which the different galaxies are observed. We assigned an $R_c$ value to each 
galaxy as the radius where its brightness profile drops below the exponential fit. 
     
\begin{figure}[!t]
\plotone{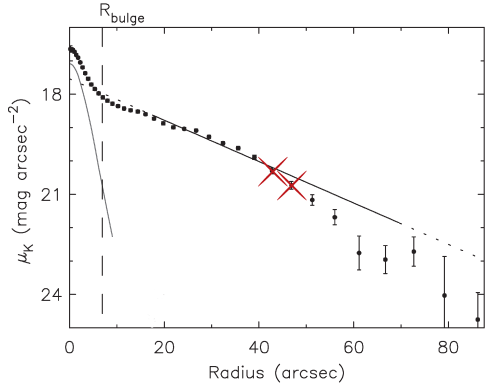}
\caption{The figure shows how we take the critical radius, $R_{c}$ of each galaxy, from their K-band surface brightness profiles,
here is shown an example for $UGC 4036$, the profile falls bellow the exponential fit between the points marks with crosses the midpoint is
$R_c=44.8 arcsec$ the critical radius for the galaxy.}
\end{figure}

As an example of how we take the critical radius $R_c$ in each galaxy we reproduce the K-band surface brightness profile for galaxy 
$UGC 4036$ as it appears in Martinsson et al.(2013a) in figure (1), from the figure we can see that the profile falls below the exponential 
fit between the points marked with crosses, we take the midpoint $R_c=44.8 arcsec$ as the critical radius for $UGC 4036$, The error 
associated to $R_c$ is the radial interval between the points where the fall occurs. We proceed in the same way for all galaxies and use 
only those where the drop is evident, which leaves us with a subsample of twenty disk galaxies.

\begin{figure}[!t]
\plotone{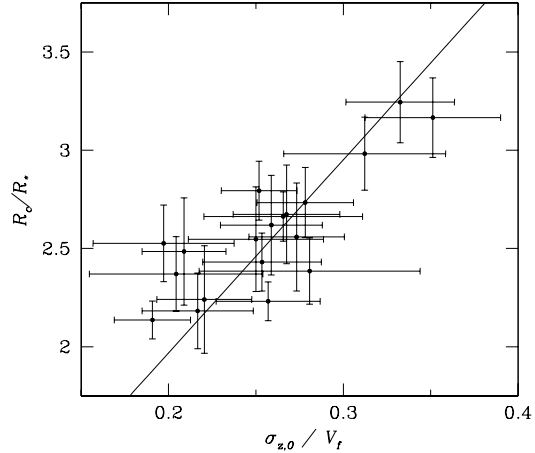}
\caption{The figure shows the critical radius where the light profile of each galaxy drops below the exponential fit, $R_{c}$,
in units of the disk light disk scale length, $R_{\star}$, of each system, as a function of the ratio of the central
disk vertical velocity dispersion, $\sigma_{z,0}$ to the flat rotation velocity amplitude of each galaxy. The solid line 
gives the best fit direct proportionality between the two quantities, a result expected from the MONDian disk stability
analogue to the classic Toomre stability criteria, and which can be seen to be an excellent fit to the observational data.}
\end{figure}

In figure (2) we plot $\frac{R_{c}}{R_{\star}}$ and $\frac{\sigma_{0}}{V_{f}}$ for the galaxies in our sample. It can be seen that 
the result is consistent with a straight line, as predicted by eq.(13), with $\beta= 9.84 \pm 0.24$. After assigning an error to 
$R_{c}$ as described above, we perform the full error propagation to calculate the corresponding $1\sigma$ confidence intervals 
for the quantities $\frac{R_{c}}{R_{\star}}$ and $\frac{\sigma_{0}}{V_{f}}$ using the errors reported in Martinsson et al.(2013a) 
for $V_{f}$, $R_{\star}$ and $\sigma_{z,0}$ to arrive at the error bars show in the figure.

We note that for our sample $\langle \frac{R_{\star}}{R_{\sigma z}} \rangle = 0.43$ with a dispersion of 0.21, supporting the  
assumption used to derive the eq. (13) of $R_{\sigma} \approx 2R_{\star}$. $UGC 4455$ was not considered for our study, as it is
the only galaxy in Martinsson et al. (2013a) described by the authors as having strong spiral arms which might disrupt the determination
of the surface brightness profile. Asides from $UGC 4455$, no galaxy with a clearly evident dip below the exponential 
profile was excluded from our sample, the 10 excluded ones show no evident dip at the radii predicted by eq.(13), or at any other one.
This probably reflects the sensitivity of K-band observations to the integral star formation history of a galaxy, if changes in
surface density, stellar and gas mass profiles, and velocity dispersion profiles (or amplitude of the rotation velocity curve even)
have shifted $R_{c}$ significantly during the course of evolution, no trace of this feature would be imprinted upon the present K-band
surface brightness profile. A better test of the ideas here presented would require a large homogeneous sample including an indication
of the ionisation state of the gas along the disk.

A more detailed calibration of the numerical proportionality factor at the critical radii can be obtained from a further subset of 
galaxies where all the relevant quantities are directly available at $R_{c}$, and no global fits are used. For these we obtain 
$\langle  (\sigma /\Omega)^{2}/(a_{0}G \Sigma)  \rangle = 0.4$ at the critical radii. Notice that this last number can not be directly 
compared to the previous, more approximate calibration, of $\beta=9.84$, as passing from eq.(8) to eq.(13) implies assuming  
$M=2\pi R^{2}_{\star} \Sigma_{0}$, where $M$ is the total baryonic mass of the galaxy, something that is not exactly accurate, to a degree
which depends on the bulge fraction, gas fraction, and details of the mass to light ratios (and their probable radial variations) used in 
estimating the surface density mass profile from the observed surface brightness one. Notice also that if one is to asses the validity
of a modified gravity law, $\Sigma$ estimates which rely on classical gravity virial relations for the vertical disk structure are not
relevant. Using the same data above to evaluate the classical $\langle Q \rangle = \langle  (\sigma /\Omega)/(G \Sigma)  \rangle$
at the critical radii gives a value of  $\langle Q \rangle = 4$, and hence the disks studied here appear classically as the low surface 
density 'sub-critical' hot, large $Q$ systems of Martin \& Kennicutt (2001).

We end this section with Table (1), which summarises the galaxies and parameters used in this study.

\begin{table*}
\begin{flushleft}
  \caption{Parameters for the galaxies treated.}
  \begin{tabular}{@{}lllll@{}}
  \hline
 \hline
   UGC    &   $V_{f} (km/s) $ & $R_{\star} (kpc)$  & $\sigma_{z,0} (km/s)$ & $R_{c} (kpc)$\\
 \hline
 448      & $186.00\pm 11.00$   & $3.86\pm 0.21$    & $47.8 \pm 2.70$   & $8.61 \pm 0.47$ \\
 463     & $209.00 \pm 12.00$  & $3.78\pm 0.27$   & $69.5 \pm 2.50$   & $12.26 \pm 0.77$  \\
 1081      & $156.00\pm 9.00$   & $3.05\pm 0.28$    & $43.40 \pm 1.80$  & $8.35 \pm 0.62$  \\
 1087      & $160.00 \pm 1.00$  & $3.23 \pm 0.21$   & $42.50 \pm 4.60$  & $8.61\pm 0.45$  \\
 1635      & $152.00\pm 9.00$   & $2.92 \pm 0.21$   & $29.00  \pm 1.60$  & $6.24 \pm 0.42$  \\
 1862      & $102.00\pm 8.00$   & $1.40 \pm 0.39$ & $26.40 \pm 0.90$   & $3.68\pm 0.51$  \\
 1908      & $237.00 \pm 14.00$   & $4.86 \pm 0.19$ & $74.00 \pm 6.60$   & $14.51\pm 0.70$  \\
 3091      & $156 \pm 10.00$   & $3.60 \pm 0.34$ & $33.80  \pm 2.80$  & $7.85 \pm 0.57$  \\
 3140     & $209 \pm 12.00$   & $3.51 \pm 0.28$ & $73.40  \pm 3.90$  & $11.12 \pm 0.61$ \\
 3701     & $124.00 \pm 9.00$   & $3.55 \pm 0.60$ & $25.90  \pm 1.10$  & $8.83 \pm 0.57$  \\
 3997      & $154.00 \pm 11.00$   & $5.54 \pm 0.43$ & $38.50  \pm 3.20$  & $14.10 \pm 0.59$  \\ 
 4036      & $187.00 \pm 11.00$   & $4.32 \pm 0.71$ & $51.10  \pm 2.10$  & $11.06 \pm 0.75$  \\
 4107     & $166.00 \pm 10.00$   & $3.20 \pm 0.29$ & $41.80  \pm 1.07$  & $8.94 \pm 0.53$  \\
 4368    & $163.00 \pm 10.00$   & $3.19 \pm 0.35$ & $43.60  \pm 2.30$  & $8.53 \pm 0.48$  \\
 4622     & $229.00 \pm 14.00$   & $7.56 \pm 0.28$ & $46.80  \pm 8.50$  & $17.92 \pm 0.51$  \\
 6903      & $143.00 \pm 10.00$   & $4.22 \pm 0.83$ & $28.20  \pm 3.80$  & $10.67 \pm 1.03$  \\
 7244     & $132.00 \pm 9.00$   & $3.86 \pm 0.70$ & $29.10  \pm 1.60$  & $8.66 \pm 0.48$  \\
 7917      & $249 \pm 14.00$   & $8.46 \pm 0.39$ & $69.90  \pm 11.80$  & $20.18 \pm 0.94$  \\
 12391     & $172.00 \pm 10.00$   & $3.86 \pm 0.54$ & $43.60  \pm 3.30$  & $9.39 \pm 0.46$  \\
\hline
\end{tabular}

List of galaxies from Martinsson et al. (2013a) and Martinsson et al. (2013b) used. The first three properties
are the reported observational estimates from the above authors, and $R_{c}$ gives our estimate from the radial
K-band surface brightness profiles. Galaxies UGC 00448, UGC 01635, UGC 03091, UGC 04036, UGC 04368 and UGC 12391 
have well measured values for both disk velocity dispersion and mass surface density estimates at $R_{c}$, and 
were used in the estimates of $\langle Q_{M} \rangle$ and $\langle Q \rangle$ at the critical radii given at the 
end of section (3).
\end{flushleft}
\end{table*}

\section{Conclusions}
We develop under MONDian gravity the equivalent to Toomre's first order disk stability criterion, by
analogy to the classical case, through a comparison of the tidal critical density and a Jeans critical
density. Our result is hence a local criterion referring to the ability of a 3D disk element to undergo
collapse under its self-gravity. 

The resulting critical parameter can be found to be $Q_{M}=(\sigma \Omega/a_{0}) Q=(\sigma \Omega)^{2}/(a_{0} G \Sigma)$.

A comparison to the critical radii at which the observed surface light profile drops below the exponential fit, in a sample of 
disk galaxies with measured rotation curves and velocity dispersion profiles, shows much better agreement with a 
$Q_{M}$ criterion than with a classic $Q$ one.

It appears that the physics of tides and internal dynamical pressure balancing self-gravity in disk galaxies is much
better represented by MONDian physics than by classical gravity.

\section{Acknowledgements}

Xavier Hernandez acknowledges financial assistance from UNAM DGAPA grant IN100814. Alejandra Jimen\'ez acknowledges 
financial support from a CONACYT scholarship.

\end{document}